\documentclass[fleqn,10pt]{wlscirep}
\usepackage[utf8]{inputenc}
\usepackage[T1]{fontenc}
\usepackage{float}
\usepackage{subfig}
\usepackage{subcaption}
\newcommand{\merra}{MERRA-2}
\newcommand{\ibtracs}{IBTrACS}

\title{Reconstructing Pre-Satellite Tropical Cyclogenesis Climatology Using Deep Learning}
\author[1,*]{Chanh Kieu}
\author[2]{Thanh T.N. Nguyen}
\author[2]{Duc-Trong Le}
\author[2]{Duc Gia-Anh Hoang}
\author[3]{Quang-Lap Luu}
\author[2]{Binh T. Dang}
\author[2]{Truong X. Ngo}
\author[4]{Quang-Trung Luu}
\author[5]{Tien D. Du}
\author[5]{Khiem V. Mai}
\affil[1]{Department of Earth and Atmospheric Sciences, Indiana University, Bloomington, Indiana 47405, USA}
\affil[2]{Faculty of Information Technology, VNU University of Engineering and Technology, Hanoi, Vietnam}
\affil[3]{School of Electrical and Electronic Engineering, Hanoi University of Science and Technology, Hanoi, Vietnam }
\affil[4]{Universit\'e Paris-Saclay - CNRS - CentraleSup\'elec - L2S, Gif-sur-Yvette, F-91192, France}
\affil[5]{Vietnam National Center for Hydro-Meteorological Forecasting, Hanoi, Vietnam}
\affil[*]{Corresponding author: ckieu@iu.edu}

\begin{abstract}
A reliable tropical cyclone (TC) climatology is the key to assessing historical and future changes in TC activities. While global TC records have been systematically maintained since the early 1940s, substantial uncertainties remain for the pre-satellite era during which TC observations relied mostly on scattered aircraft reconnaissance and sporadic ship reports. This study presents a deep learning (DL) approach to reconstruct historical TC activity in the western North Pacific (WNP) basin, with a main focus on the pre-satellite era. Using data feature enrichment tailored for tropical cyclogenesis (TCG), we demonstrate that DL can effectively capture the main characteristics and changes in TCG climatology during the post-satellite era. With additional cross-validations, the reconstruction of TCG climatology is then extended to a pre-satellite period (1940-1960) during which TC base-track datasets are most uncertain. Our DL reconstruction reveals a significant missing of TCG in the current best-track data between September and November during the pre-satellite era. Such a TCG undercount in the best track data occurs mainly around 10-15$^\circ$N in the central WNP, while coastal regions show better consistency with DL reconstruction. These findings not only highlight the potential of DL for improving historical assessments of TC activity, but also advance our understanding of TCG processes by identifying key environmental conditions conducive to TC formation. The DL approach presented herein can be applied to other ocean basins, climate proxies, or reanalysis datasets for future TC climate studies.
\end{abstract}
\begin{document}
\flushbottom
\maketitle

\section*{Introduction}
Climatologically, any observable shifts in tropical cyclone (TC) characteristics such as frequency, genesis locations, or intensity, are often considered as manifestations of large-scale climate variability. Due to the limited length and quality of historical TC records, current climate assessments or projections for TC activities carry, however, considerable spatial and temporal uncertainties, especially when examining TC activities over a sufficiently long period of time \cite{Knutson_etal1998,  Bengtsson_etal2007, Cha_etal2020, Kieu_etal2023, Kieu_etal2025}. 

While modern TC best-track datasets such as the International Best Track Archive for Climate Stewardship (IBTrACS, \cite{IBTRACS}) provide records extending back to the early 1840s for some ocean basins, these historical data have been mostly under-utilized in TC climate studies due to concerns about their reliability \cite{Landsea_etal2006}. Prior to the availability of satellite data in the late 1970s, TC monitoring relied primarily on aircraft reconnaissance, ship reports, and inland surface observations. These methods have a number of limitations such as the potential omission of TCs over the open ocean, infrequent reconnaissance flights, inconsistencies in estimating surface winds from different flight-level data, or data quality issues \cite{UhlhornBlack2003, Franklin_etal2003, Delgado_etal2018, Sheets1990, ChangGuo2007, Rappaport_etal2009}. As a result, assessing TC climate variability during the pre-satellite era (i.e., prior to 1980) remains challenging and often not included in TC climate research.

At the basin-wide level, the quality of TC records is even more variable. For example, TC observations in the western North Pacific (WNP) could date back to 1884 \cite{IBTRACS}, yet these records have undergone continual revisions due to advancements in observation systems, new sources of information, and retrieval methodologies \cite{Chu_etal2002, NakazawaHoshino2009}. Although aircraft reconnaissance in the WNP basin became available in 1948, these early data primarily covered the southwestern portion of the WNP basin and were limited by sparse surface wind observations and varying flight-level wind estimates \cite{Hagen_etal2012, Delgado_etal2018}. The availability of geostationary satellite and the introduction of the Dvorak technique emerged only after 1975 \cite{Dvorak1975, Dvorak1984} while the routine aircraft reconnaissance in the WNP by the US ceased by the mid-1980s, making it hard to rely on any TC data prior to 1980 for any climate change study in the WNP basin.


Given those inherent issues with the quality of TC records, climate change analyses or assessments based on any best-track TC records have generally been restricted to the post-satellite era, when data homogeneity is more reliable \cite{ipcc5, Emanuel2005, Moon_etal2019, DownsKieu2020}. This constraint on the temporal extent of TC climate research in fact holds true across ocean basins, albeit the exact starting years may slightly vary among the North Atlantic, WNP, Indian Ocean, or eastern Pacific basin. Having longer and more reliable TC records is, therefore, desirable for studying historical and future changes of TC climatology. In this regard, any approach or techniques capable of reconstructing TC climatology further back in time would be valuable, not only for assessing past climate variability but also for improving future projections or enhancing seasonal to decadal outlooks of TC activities.      

The rapid advancement of machine learning (ML) techniques has opened up new directions for their application across domains in atmospheric science. With the vast amount of observational and model-generated data, weather and climate research naturally align with the big data paradigm, offering unique opportunities to train ML models for both short-term weather forecasting and long-term climate research \cite{Schultz_etal2021, pathak_etal2022, Remi_etal2023, NguyenKieu2024}. In fact, various ML architectures have been recently proposed, which outperform traditional physics-based models ifor short-range weather prediction \cite{pathak_etal2022, Bi_etal2021, Remi_etal2023}. 

Specific to TC research, most ML applications have focused so far on short-term weather forecasting, satellite data retrieval, or diagnostic analysis. For example, a range of deep-learning (DL) models that use satellite data to train convolutional neural network models capable of categorizing TC intensity based on cloud patterns have been proposed and demonstrated some degree of success \cite{miller_etal2017, Wimmers_etal2019, NguyenKieu2024, Mu_etal2024, Tong_etal2025, Le_etal2025}. Other studies have enhanced weather nowcasting and diagnostic performance by incorporating TC tracking information and reanalysis data, offering some improvements for operational practice \cite{Gao_etal2018, Kim_etal2019, Giffard_etal2020}. 
While these studies addressed different aspects of TC intensity and statistics for weather prediction problems, they also suggested the potential of DL for TC climate research beyond traditional physical-based downscaling that we wish to examine in this study.  


With the growing applications of DL, the primary objective of this study is to present a DL framework for reconstructing historical TC climatology from climate reanalysis datasets. Although our ultimate goal is to extend TC records as far back into the pre-satellite era as possible for all ocean basins, we will focus specifically on TCG climatology in the WNP basin in this study. Building upon our recent DL models for TCG (Le et al. (2025) \cite{Le_etal2025}, hereafter L25), we employ DL to reconstruct historical TC climatology for several different pre- and post-satellite periods, using the NASA Modern-Era Retrospective analysis for Research and Applications, version 2 (MERRA-2), and the ECMWF Reanalysis, version 5 (ERA5, \cite{ERA5}). By comparing these results with existing best-track database (\ibtracs) during the pre- and post-satellite eras, we can assess both the efficiency of our DL approach and the uncertainty of the current best-track database in the past. From a climate downscaling perspective, our DL-based climatology thus offers a new validation tool complement to traditional vortex-tracking methods, highlighting the significance as well as the broader potential of DL in TC climate studies.

%
%
\section*{Results}
TCs are multi-scale systems that exhibit different weather and climate characteristics. As such, their analyses generally encompass a wide range of metrics such as intensity, frequency, accumulated cyclone energy, track, occurrence, genesis potential indices, lifetime maximum intensity, among many others. Any effort to study or reconstruct TC climatology must therefore be specifically tailored for a given metric to ensure that the associated models or datasets can be optimized accordingly \cite{Emanuel2005, Camargo_etal2007a, Moon_etal2019, Cha_etal2020, Vu_etal2024, Kieu_etal2025a, Kieu_etal2025}.

In this study, two key metrics of TCG climatology will be presented for the WNP basin, which include ($i$) the spatial distribution of TCG density and ($ii$) the seasonality of TCG frequency. 
With favorable environmental conditions for TCG such as warm sea surface temperatures (SSTs), active monsoon troughs, and frequent convectively coupled equatorial waves, the WNP is statistically the most active ocean basin for TC activities, producing approximately 25-32 TCs annually. As a result, changes in TC climatology often exhibit stronger and more discernible signals in this region, making the WNP an ideal “testing ground” for studying TC climate variability \cite{Kossin_etal2016, Peduzzi_etal2012, DefforgeMerlis2017, Duong_etal2021, Thanh_etal2020}. 
These TCG metrics serve as an early testament of our DL approach that we will expand further to other TC metrics and ocean basins in the future as outlined later in the Discussion section.    

\subsection*{Post-satellite TC climatology reconstruction} 
\begin{figure}[ht]
\centering
\includegraphics[width=16.2cm]{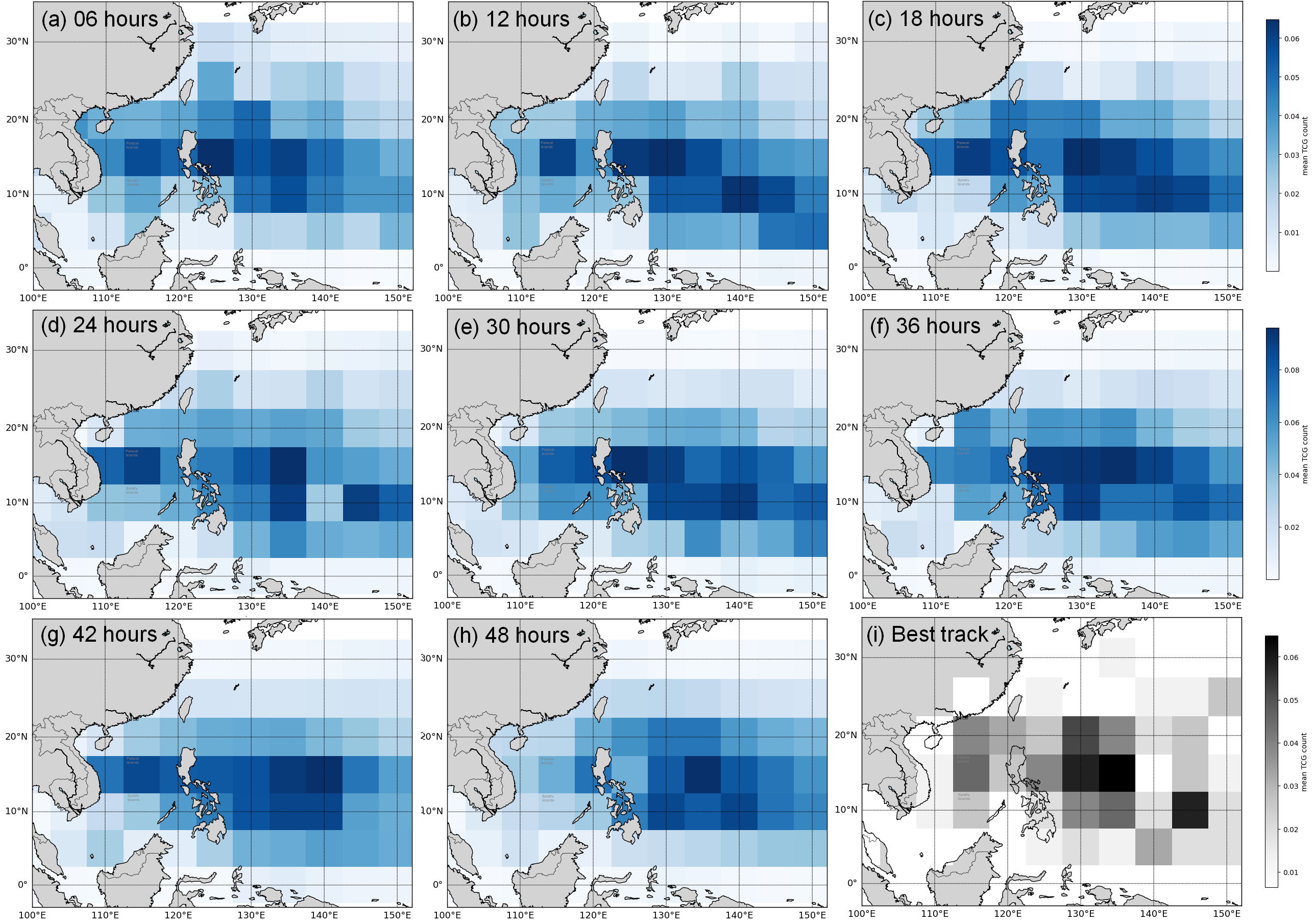}
\caption{(a)-(h) Spatial distribution of five-year average of TCG probability predictions (blue shading) from the ResNet-18 DL model during the test period from 2017--2022, based on different data enrichment windows ranging from 6 to 48 hours prior to the first moment of a Tropical Depression record in the best track data; and (i) observed TCG density (gray shading) for the same 2017--2022 period that is derived from the best-track data. Here, the observed TCG density is defined as the total number of actual TCG events reported in each grid box during the test period, divided by the total number of TCG events recorded in the best-track dataset for the same test period. Note that different shading scales are used for the best-track data and DL predictions to better display the contrast in areas of maximum TCG probability in the DL model.}
\label{fig:dist_1722}
\end{figure}

To have a broad perspective of how our DL model reconstructs TCG climatology during the post-satellite area as compared to the observational data, Fig. \ref{fig:dist_1722} shows the spatial distributions of TCG density for the test period 2017--2022, with a range of data enrichment windows described in the Method section. Here, the model was trained with the MERRA-2 and ERA5 data during 1980--2016, using the same DL architecture based on the residual convolutional neural network (ResNet-18) described in the Method section. Following L25's study, we use this DL model to reconstruct TCG climatology for non-overlapping $5^\circ \times 5^\circ$ grid boxes over the WNP basin by calculating the mean TCG probability for each box during the test period. 

%
%



One can see from Fig. \ref{fig:dist_1722} that the DL model produces a TCG density distribution broadly consistent with observations across a range of data-enrichment windows, even when using climate input data at 0.5$^\circ$ resolution. The consistency across data enrichment windows is important for our TCG problem herein, because it demonstrates that some potential environmental signals inherent for TCG actually exist for some time period and are sufficiently well-defined that a DL model can learn and capture. From a practical perspective, this window is necessary, as the exact timing of TCG reported in the best-track dataset differs from the first time a TC is recorded in the best track and is often highly uncertain. Thus, one should expect a window containing the right TCG environments rather than a single moment of TCG for DL models to learn. As shown in Fig. \ref{fig:dist_1722}, the favorable environments for TCG within an optimal window as obtained from our DL enrichment indeed support the fact that most DL-reconstructed TCG events cluster along a monsoon trough in the WNP basin. Of note, the dominant TCG density maxima over the western Philippine Sea (135$^\circ$–145$^\circ$E) are also captured by our DL reconstruction, indicating the ability of the DL model in learning the right environmental signals for TCG development. 

With this capability, our DL-based reconstruction can reproduce the observed TCG climatology with spatial correlations ranging from 0.75 to 0.81, depending on data enrichment windows. Among these windows, the model performs best for enrichment periods between 6 and 30 hours, achieving a maximum correlation of 0.81 for the 18-hour window. For longer windows, the model performance starts to decline, because excessive past data used to enrich TCG information obscures the relevant environmental signals needed for TCG detection. As a result, incorporating data from more than 36 hours prior does not improve the reconstruction and instead leads to a reduced TCG detection probability and a higher false-alarm rate, consistent with the findings of L25.

%
%
\begin{figure}[ht!]
\begin{center}
\includegraphics[width=12cm]{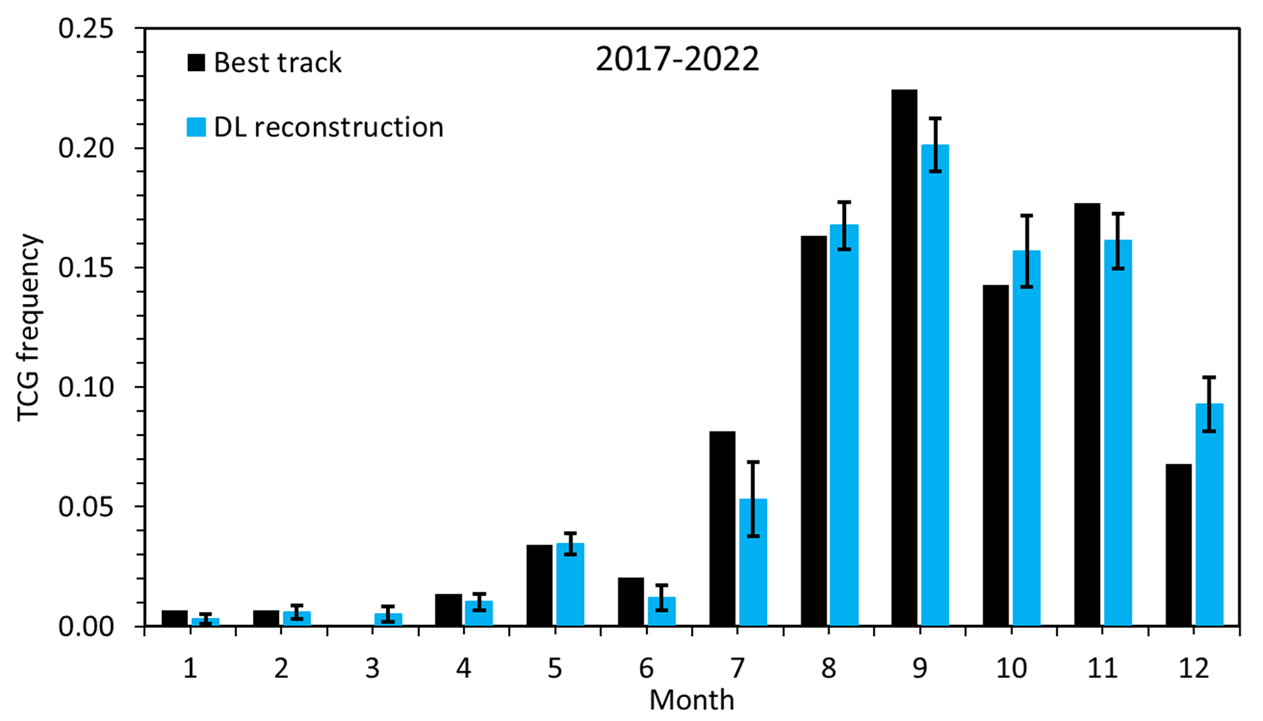}
\caption{Monthly distribution of TCG frequency detected in the WNP basin from the test data (2017--2022), using the ResNet-18 model with 8 different time windows from 6 to 48 hours (blue columns). Black columns denote the actual TCG frequency obtained from the best-track database \ibtracs\ during the same period. The error bars represent the statistics of DL predictions using different data enrichment windows.}
\label{fig:dist_monthly}
\end{center}
\end{figure}

Looking at the seasonality of TCG frequency (Fig. \ref{fig:dist_monthly}), one further notices that DL reconstruction can also recover the monthly distribution of TCG frequency, with the highest TCG frequency in July-August consistent with the observed distribution derived from the best-track data. Of interest is that the DL model appears to capture the second peak of TCG frequency in October, albeit the double-peak pattern obtained from our DL model is less pronounced compared to the observed TCG frequency.

Such a good performance of our DL model in reconstructing TCG density and seasonality is noteworthy if one recalls that the model learns the pattern of TCG purely from training data during 1980--2016, while the test data spans 2017--2022. This non-overlapping of test and training data indicates that TCG processes as well as their climate characteristics can be learned from the coarse-resolution environments in one period and generalized to the other, once a DL model is properly designed. This ability is especially important in the context of climate change, because any two different climate regimes will likely share some common climate modes as well as possible shifts between the two. Being able to learn the common variability characteristics is, therefore, a must for any DL model before it can be used to detect any climate change signal.      

As with traditional vortex tracking algorithms in physical-based models, we note that any DL model used for reconstructing TC climatology will inherently face some metric-dependent limitations. That is, optimizing a DL model for one metric may cause some bias in other metrics, which explains some discrepancies in both TCG density and seasonal distribution (Figs. \ref{fig:dist_1722}-\ref{fig:dist_monthly}) as compared to the observation. Although we can always improve one metric over the other by tuning model hyperparameters, it is generally difficult to optimize a model for all metrics simultaneously, regardless of the downscaling or DL methods \cite{Vu_etal2024, Kieu_etal2025}. Despite these discrepancies, the overall consistency between the DL-reconstructed and observed TCG distributions as shown in Figs. \ref{fig:dist_1722}-\ref{fig:dist_monthly} provides us with confidence in using our model to explore TCG climatology and its related variability, as we now turn to in the next section.

\subsection*{Post-satellite climatic shift}
Building on the capabilities of the DL model established in the previous section, we examine next some potential changes in TCG climatology between two post-satellite era periods 1980–1985 and 2017–2022. Here, the 1980–1985 period is chosen because it includes one of the strongest El Ni\~no Southern Oscillation (ENSO) events on record \cite{McPhaden_etal2006, Zhang_etal2019}, providing a valuable opportunity to evaluate the model's ability in capturing the variations under abnormal conditions. This analysis thus provides an additional validation for our DL model in reproducing observed changes in TCG climatology during the post-satellite era and offers more insights into its generalization capability. 
%
%
\begin{figure}[ht!]
\centering
\includegraphics[width=16.2cm]{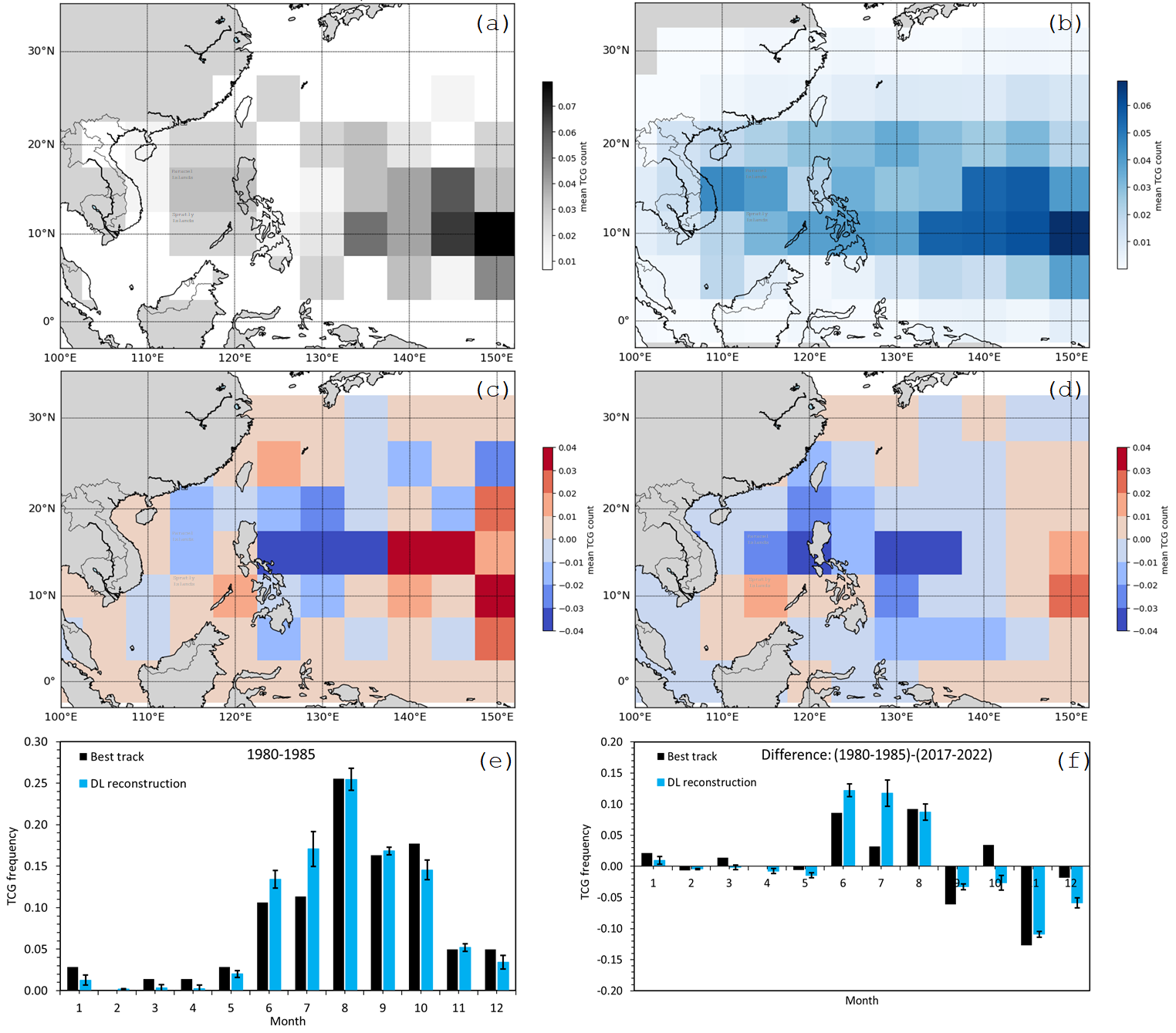}
\caption{(a)-(b) Similar to Fig. \ref{fig:dist_1722}, but using the 1980--1985 period as the test data for the best-track TCG density (gray shading, left panel) and the DL-reconstructed TCG density using a 18-hour data enrichment window (blue shading, right panel), (c)-(d) the difference between the 1980-1985 period and the 2017--2022 period, and (e) similar to Fig. \ref{fig:dist_monthly} but for the TCG seasonality during the 1980--1985 period (left panel) as obtained from the best-track data (black columns) and DL reconstruction (blue columns), and (f) the relative difference between the 1980-1985 and the 2017-2022 periods (right panel).}
\label{fig:dist_8085}
\end{figure}

In this regard, Fig. \ref{fig:dist_8085}a–b presents the observed TCG density distribution for the 1980--1985 period alongside the DL reconstruction, using the optimal 18-hour data enrichment window that yields the highest spatial correlation. Looking at the observed TCG density (Fig. \ref{fig:dist_8085}a), there is a clear difference as compared to the 2017--2022 period, with a noticeable eastward shift in TCG activity toward the far eastern WNP basin during 1980--1985 (see Fig. \ref{fig:dist_8085}c). Additionally, the strong TCG activity inside the South China Sea (SCS) as seen during the 2017--2022 period is no longer apparent in 1980--1985. These substantial changes in TCG density from the best-track data highlight the strong influence of ENSO on TC development, thus providing a valuable reference for validating the performance of our DL model.

For the DL-reconstructed TCG density and its relative difference from 2017-2022 (Fig. \ref{fig:dist_8085}b,d), one can see that the changes in TCG density between the two post-satellite periods are again well captured by our DL model, especially the major shifts in the eastern Philippines Sea (between 120--140$^\circ$W) and the eastern boundary of the WNP basin (145--155$^\circ$W. See Fig. \ref{fig:dist_8085}d). The fact that the DL model can capture these important shifts is significant, as we recall that the 1980-1985 period is treated as a test period in this analysis, which is unseen to the model during the training. Thus, this reiterates that our DL model is able to correctly learn the required large-scale environments for TCG as well as the shifts in these environments between the two post-satellite periods, even when the training data does not cover the extreme ENSO mode during 1980--1985.

We note that these shifts in TCG density between the two post-satellite periods as obtained from the DL model is not limited to just one data enrichment window in Fig. \ref{fig:dist_8085}. In fact, these shifts are well captured for a range of windows from 6 to 42 hours, albeit the magnitude of the shift and spatial correlation with the observed TCG density vary with individual enrichment windows between 0.71--0.83 (see Fig. S1, Supporting Information). Unlike the 2017--2022 period, we do observe however that our DL model experiences larger variability in terms of TCG density for several enrichment windows, mainly for the 24--48 hours (see Supporting Information). This variability is most evident in the local TCG cluster over the SCS, where the DL-reconstructed TCG density is more pronounced than observed during the 1980-1985 period (cf. Fig. \ref{fig:dist_8085}b). 

Our attempt to use different DL architectures, model hyperparameter tunings, or data labeling strategies could not reduce this sensitivity for the 1980--1985 period. This suggests that the larger variability of TCG distribution during this period likely stems from the inherent limitations of learning TCG patterns from recent climate data when applied to unseen periods governed by extreme climate modes. Note again that 1980–1985 include one of the strongest ENSO events on record \cite{McPhaden_etal2006, Zhang_etal2019}, a condition absent from all training data. Because of this, the DL model’s ability to generalize to extreme periods appears constrained, unless it is trained with the full spectrum of climate variability, a scenario that is difficult to achieve in practice at present due to limited data length. This limitation becomes most apparent when the test period contains some distinct climatic anomalies, as exemplified by the 1980–1985 period.

For the change in the seasonality of TCG frequency between the two periods, our DL model demonstrates its overall ability in capturing both the peaks in TCG frequency and the shift between the two periods (Fig. \ref{fig:dist_8085}e-f). Specifically, the 2017--2022 period shows a peak in TCG frequency during August-December, while the 1980-1985 period shows the main season taking place during June-October with two small breaks in July and September (see the black columns in Fig. \ref{fig:dist_8085}c). While the DL model could not fully capture the magnitude of observed TCG frequency for every month, it does show a very consistent shift in TCG frequency between the two periods, with more TCG during June-August but less TCG during September-December as obtained from the best-track data for the 1980-1985 period. This result once again confirms the capability of our DL model in extracting meaningful TCG-related signals from large-scale environments, which are needed for later reconstructing TCG climatology further back in time.

Of course, the DL-reconstructed seasonality for the 1980–1985 period exhibits the same limitation as the TCG density in the sense that its reconstruction quality is somewhat lower than that for the 2017–2022 period, particularly in terms of the magnitude of TCG frequency. This discrepancy is most evident during June–July, when the DL reconstruction tends to overestimate the observed TCG frequency. Such inconsistencies highlight the broader challenges faced by DL models in reconstructing TCG climatology during extreme ENSO periods as mentioned above, thus emphasizing the caution when applying or interpreting DL-based results for extreme climate modes.    

We should mention that our above post-satellite analyses are limited to the choice of two 5-year periods instead of a longer period, such as 20-year periods often used in traditional climate studies \cite{ipcc5, Tran_etal2020}. This limitation is due merely to the requirement for the reliability of the DL model training, with a limited post-satellite data length from 1980 to 2020. That is, using a longer period for the test dataset would lead to a smaller training dataset, and so the model output would no longer be robust. With more data in the future, our model and approach presented herein can reach better results as designed.

Despite some issues in the exact magnitude of TCG density or frequency under extreme ENSO mode during 1980--1985, the broad consistency in both the characteristics and the shift of TCG distribution between the two post-satellite periods suggests that DL can indeed learn and generalize key properties of large-scale environments for TCG climatology. The performance of our specific DL design obtained herein demonstrates the capability of DL in reconstructing not only TCG climatology but also its climatic changes, which will become more effective when a longer training dataset is available in the future. 
%
%
\begin{figure}[ht]
\begin{center}
\includegraphics[width=16.2cm]{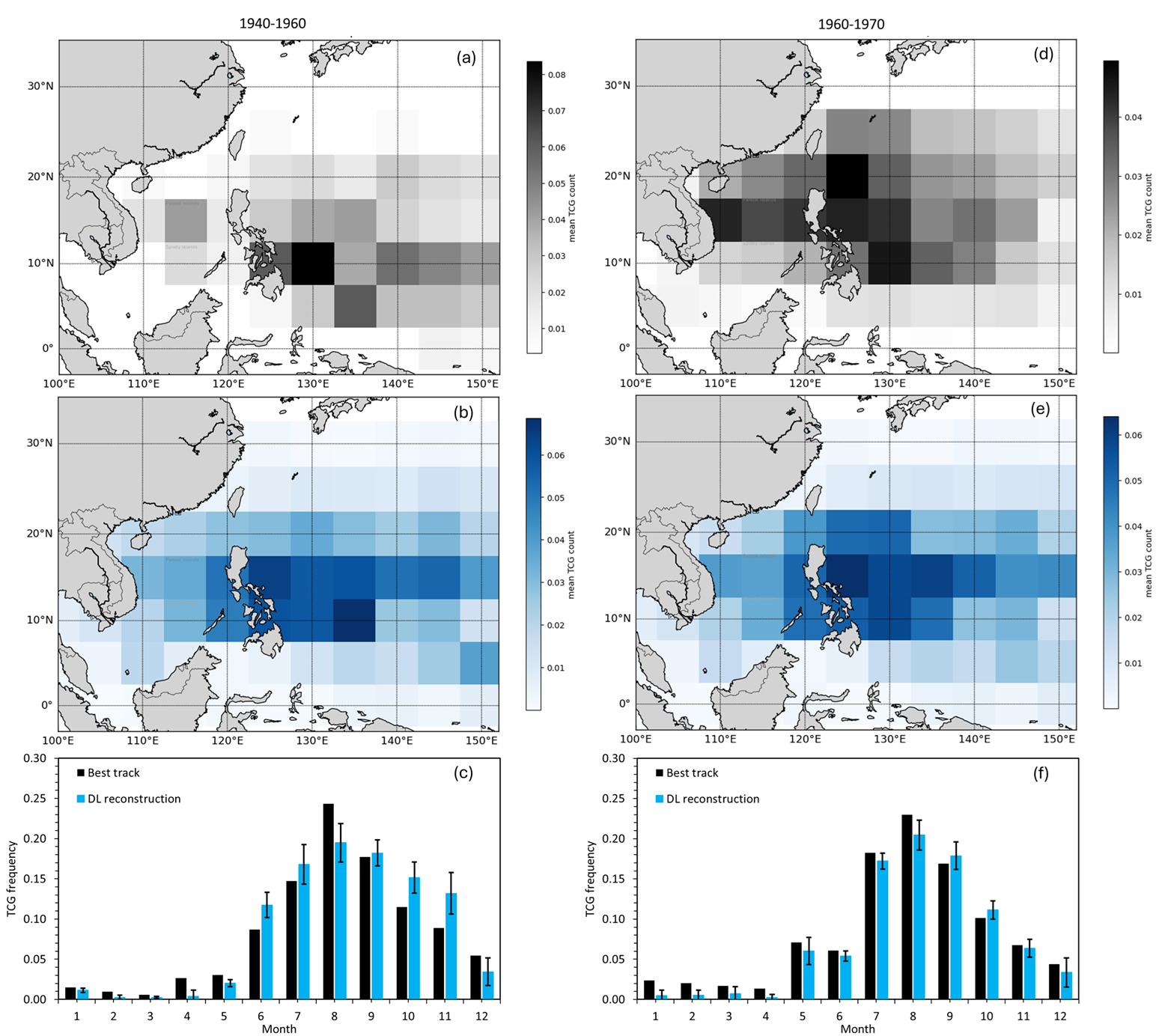}
\caption{Similar to Fig. \ref{fig:dist_8085}a,b,e but for the TCG density (shaded) and frequency (columns) distribution during the pre-satellite 1940--1960 period, using the training data from MERRA-2 and ERA5 during 1980--2020. (d)-(f). Similar to (a)-(c) but for the pre-satellite 1960-1970 period.}
\label{fig:dist_4060}
\end{center}
\end{figure}

\subsection*{Pre-satellite TC climatology reconstruction}
With the verified performance and optimization during the post-satellite period as presented in the previous sections, our final step is to apply the DL model to reconstruct TCG climatology for a pre-satellite period during which the best-track data is most uncertain. Specifically for this purpose, we will choose the farthest possible period from 1940--1960 for which both modern reanalysis data and TC best-track data are available, such that the reconstructed TCG climatology can be compared to that from the best-track data. 

Note that for this 1940--1960 period, MERRA-2 data is unavailable. Therefore, only the ERA5 dataset is used for TCG reconstruction. In addition, we use the entire 1980-2020 data from MERRA-2 for pre-training our model and then fine-tune it with the ERA5 to increase the data length before extrapolating back into 1940-1960. This approach targets our broader aim of refining a DL model trained on one dataset with another dataset for future applications, a process known as transfer learning or fine-tuning that is often used to help improve model performance. In addition, the 20-year period for this pre-satellite reconstruction also provides more robust statistics, as compared to the 5-year reconstruction presented in the post-satellite analyses. Thus, this period allows us to better demonstrate the contrasts between the existing best-track database and our DL reconstruction during the pre-satellite era as expected.

Figure \ref{fig:dist_4060} compares the TCG density and frequency derived from the best-track data with those from DL reconstruction for 1940--1960. In general, there is some agreement in terms of the clusters of TCG density between DL reconstruction and the best-track data, mostly over the central WNP basin and the SCS. The consistency observed in the SCS is of interest, considering that prior to the satellite era TC observations were primarily limited to landfalling storms or systems near coastal areas or shipping routes. This agreement between our DL reconstruction and best-track data in the SCS therefore indicates that there is a good degree of reliability of the best-track data in the SCS region.

In the open ocean near the eastern Philippine Sea, several noteworthy features emerge. First, the DL model could identify the main TCG cluster in the central WNP basin (Fig. \ref{fig:dist_4060}b) somewhat similar to the TCG density derived from the best-track dataset. However, the location of the maximum TCG density in the DL reconstruction is at higher latitudes (10-15$^\circ$N), whereas it is concentrated further south (around 10$^\circ$N) in the best-track data. Consequently, the spatial correlation between DL-reconstructed and the best-track TCG density now drops to 0.54-0.77, which is significantly lower than those obtained during the post-satellite period. This substantial discrepancy between the best-track dataset and DL reconstruction can be seen in all data enrichment windows (see Supporting Information), thus giving us more confidence in such a gap between the DL reconstruction and the best-track data during 1940-1960. 

Second, the seasonality of TCG frequency in Fig. \ref{fig:dist_4060}c provides additional information into the months during which the discrepancy between the best-track data and the DL reconstruction is most realized. While sharing a broadly similar seasonal distribution with the peak of TCG frequency during June-November, the best-track data reports lower TCG frequencies compared to the DL reconstruction (except for August). If one trusts the DL reconstruction based on its performance during the post-satellite era, this means that the best-track database may have missed a significant portion of TCG events during these months. When considered alongside the spatial distribution of TCG density in Fig. \ref{fig:dist_4060}a, these results thus imply that \textit{the current best-track database \ibtracs\, likely undercounts a significant number of TCs in the central WNP between 10-15$^\circ$N latitudes during September-November of the 1940–1960 period}.  

To further confirm whether the systematic bias in TCG density and frequency during 1940–1960 is representative for other pre-satellite periods, Fig. \ref{fig:dist_4060}d-f presents the DL-reconstructed TCG climatology for 1960-1970. Here, we restrict our reconstruction to this decade to avoid overlap with the 10-year period from 1970–1980 during which some satellite observations became partially available. It is apparent from this reconstruction that the bias in TCG frequency during September–November remains evident, thus providing additional confidence in the undercount of TCGs during these months. 

On the other hand, the observed TCG density pattern for 1960-1970 does not exhibit the pronounced void in the central WNP basin as seen in the 1940–1960 reconstruction. Its overall spatial distribution actually more resembles the post-satellite era, with clusters of TCG around 10–15$^\circ$N. This is intriguing, as the best-track data for 1960–1970 appear substantially different from those for 1940–1960 despite the same lack of satellite input in these two periods. In contrast, the DL reconstruction could maintain a persistent TCG distribution that is broadly consistent across both post-satellite and pre-satellite periods, which underscores the robustness of its reconstruction capability.  

In general, such differences between the DL reconstruction and the best-track dataset during the pre-satellite era, especially for the 1940-1960 period, may stem from several factors including: ($i$) limited TC observations over the open ocean, ($ii$) potential uncertainties in ERA5 reanalysis data, ($iii$) limitations in our DL model learning or generalization, and ($iv$) a combination of all these factors. While it is challenging to disentangle the relative contributions of each factor, assuming the reliability of ERA5 data extending back to the 1940s and the robustness of our DL model in reconstructing past climatology as verified during the post-satellite era, the results herein could present an approach to identify where and when TCs may be missing from the current best-track data in the WNP basin. Of course, our assumption on the quality of the ERA5 data for the pre-satellite period over any open ocean is strong. However, this assumption is justified if we note that global reanalysis data are generally more reliable than records of local extreme events during 1940-1960, given the current global assimilation advances. Therefore, having an effective way to extract extreme events from these global reanalysis datasets is more advantageous than relying on local records of extreme events for climate analyses. 

With the above DL reconstruction of historical TCG climatology, it is tempting to compare the pre- and post-satellite periods to assess the potential change of TCG climatology over time. While such comparisons can always be carried out, we note that the pre-satellite reconstruction spans a 20-year period, whereas the post-satellite reconstruction shown in Fig. \ref{fig:dist_1722} or \ref{fig:dist_8085} covers only five years. Despite this temporal mismatch, the DL reconstruction could at least provide different changes in TCG density from that drawn purely from the best-track dataset. How much confidence one can assert for these changes is nevertheless difficult to evaluate due to the absence of other independent pre-satellite TC observations over the open ocean. Therefore, we do not attempt to draw further conclusions regarding climate changes in TCG across these periods. Instead, our analyses herein focus more on the differences in TCG climatology between the best-track data and the DL reconstruction for the same historical time period as shown in Figs. \ref{fig:dist_4060}.

\section*{Discussion}
In this study, we presented a deep learning (DL) framework to reconstruct the historical climatology of tropical cyclogenesis (TCG), with a focus on the pre-satellite era during which the TC best-track database contains significant uncertainties. Reconstructing TC activity during this pre-satellite era is essential for advancing our understanding of TC climate variability, as a reliable historical climatology provides a foundation for assessing both past and future changes in TC behavior.

Specifically for the DL application to TCG climatology, we implemented a temporal data-enrichment strategy within a residual convolutional neural network architecture to address the inherent challenge posed by the limited number of TCG occurrences available for DL training. With the MERRA-2 for pre-training and ERA5 reanalysis datasets for fine-tuning, this data enrichment approach enables our DL model to more effectively learn the environmental precursors associated with TCG from reanalysis datasets at a spatial resolution of 0.5$^\circ$. Validation against two test periods during the post-satellite era showed that our DL reconstruction was able to capture key features of TCG climatology in the WNP basin, including ($i$) the clustering of TCG in the eastern Philippine Sea, ($ii$) the characteristic double peaks in seasonal TCG frequency, and ($iii$) the temporal shift of TCG patterns between two post-satellite periods.

Similar to the proxy calibration method that cross-validates two datasets over a shared period before extending it further back in time, we applied our DL model to reconstruct TCG climatology during the pre-satellite era after cross-validating it during the post-satellite era. Results showed that the main TCG clusters during 1940–1960 as reconstructed from our DL model closely resemble the modern-day TCG climatology. In particular, the main genesis cluster is located over the central WNP basin at higher latitudes instead of at lower latitudes as obtained directly from the best-track dataset. Additionally, DL reconstruction reveals more frequent TCG during September–November in 1940-1960, suggesting that a substantial number of TCs may be missing from the best-track dataset in the central WNP during these months.

Such bias in TCG frequency during the peak season is also observed for the 1960–1970 pre-satellite period. Nevertheless, the DL-reconstructed and the best-track TCG density show more consistent spatial patterns during this later period, likely because more TC observations were available with time even prior to the satellite era. Despite this high variability in the best-track TCG density, we found that our DL reconstruction still provides a coherent picture of TCG throughout the pre-satellite period. This consistency not only supports the validity of our DL reconstruction but also reinforces our confidence in its indication of potential missing TCs in the central WNP region, especially during the peak TC season of 1940–1960.


It is important to acknowledge that the analyses herein still do not fully resolve the uncertainty in TCG climatology derived from either our DL reconstruction or the best-track data. This is to some extent unavoidable due to the lack of other independent TC databases during the pre-satellite era. One potential approach to address this uncertainty is through the application of physical-based models to downscale historical climate data, which can serve as a cross-validation for the DL reconstruction. However, such downscaling efforts are themselves limited by factors such as model resolution, physical parameterizations, or challenges in detecting TC vortices, which introduce other uncertainties as well. From this standpoint, the DL reconstruction presented herein offers a complement to traditional approaches, thus providing a different way for assessing the gaps in historical TC records without requiring computationally intensive high-resolution downscaling.

The significance of our findings extends beyond the reconstruction of past TCG climatology. Specifically, our results underscore that a substantial portion of TCG processes can be effectively learned from large-scale environmental conditions. That is, \textit{reanalysis datasets contain discernible environmental signals indicative of TCG, even in the absence of explicit representations of fine-scale storm development}. Given that large-scale environmental fields are more reliably projected by climate models over decadal timescales than fine-scale extreme events, our DL approach demonstrates the potential to reconstruct TCG climatology under varying climate conditions and to assess changes in TCG across different historical periods. In this regard, DL can help identify key environmental drivers of TCG, without relying on traditional observations or physical-based modeling. 

In addition, the flexibility of our DL approach makes it promising for a wider range of future applications. For instance, the same DL framework could be applied to other historical periods or different ocean basins where the best-track data are sparse, unreliable, or unavailable, enabling the reconstruction of various TCG-related climatologies. Likewise, this method can be extended to incorporate additional metrics such as intensity, frequency, or occurrence density, allowing exploration of TC characteristics as far back in time as the availability of a reanalysis dataset could allow. As more comprehensive climate data become available, such as the 20th Century Reanalysis data \cite{Slivinski_etal2021}, that can encompass a wider range of extreme climate modes, the performance of DL-based methods is expected to improve, offering more confidence in future reconstructions and estimation of TC climate. 

Despite its promising capabilities, we also want to highlight certain limitations of the DL approach that must be addressed for future improvements. Specifically, there are still some inconsistencies in local peaks of TCG climatology within the central WNP basin, which vary depending on the choice of data enhancement windows. This issue is most evident during the 1980–1985 period, which is characterized by a very strong ENSO phase. These inconsistencies suggest possible fluctuations in reconstructed TCG density when applied to some unique periods outside the training range, thus bringing up two important issues. First, the current reanalysis datasets may still be insufficient in capturing the full range of large-scale climatic modes required for DL models to perform robustly. For example, with a limited number of ENSO cycles within 40 years of post-satellite data, DL models face challenges in learning and generalizing the influence of these climate modes on relatively rare events such as TCG. Consequently, the generalization of any DL model with any post-satellite data remains constrained, limiting its applicability to earlier periods.

Second, the difficulty of DL models in capturing all details of TCG climatology suggests that there may be some limits in quantifying or predicting TCG from just the large-scale environments \cite{Vu_etal2021, Hsieh_etal2020, NguyenKieu2024, Kieu_etal2025a, Le_etal2025}. The TCG processes have been known to involve multiscale interactions for which convective-scale processes may contribute to the timing and the probability of TCG that current TC climate datasets do not contain. How much further one can reach this predictability limit is difficult to assess, as our DL model may not yet be the best possible architecture for TCG detection or prediction. As such, this remains an open question to address before one can quantify the maximum capability for retrieving TCG information from climate datasets.       

\section*{Methods and Data}
\subsection*{Data}
In this study, two widely used climate reanalysis datasets were employed to train our DL model for reconstructing TCG climatology. The first dataset was the Modern-Era Retrospective analysis for Research and Applications, version 2 (MERRA-2), developed by the National Aeronautics and Space Administration (NASA). MERRA-2 is an atmospheric reanalysis produced using the Goddard Earth Observing System Model, version 5 (GEOS-5), along with an advanced data assimilation system. It spans the post-satellite era from 1980 to 2022 and includes improvements over the original MERRA reanalysis, such as the assimilation of newer microwave sounders, infrared radiance data, and other observational inputs.

The second dataset used was the European Reanalysis (ERA5), the fifth global atmospheric reanalysis produced by the European Centre for Medium-Range Weather Forecasts (ECMWF) \cite{ERA5}. ERA5 is available in multiple spatial resolutions \cite{Rasp_etal2020}. For consistency with MERRA-2 and to reduce computational demands, we utilized the $0.5^\circ$ latitude–longitude resolution with a 6-hour temporal interval. A subset of ERA5 variables including relative humidity, the three wind components ($u$, $v$, $w$), geopotential height, temperature, surface pressure, sea surface temperature, and landmask was selected to match those available in MERRA-2, ensuring comparability across datasets.

While both the MERRA-2 and ERA5 datasets used in this study have a relatively coarse horizontal resolution of $0.5^\circ$, which may be insufficient to fully resolve the TC inner-core structure, we deliberately adopt this resolution for two main reasons. First, using a consistent resolution across datasets facilitates the training of DL models and supports transfer learning between different reanalysis products. Maintaining the same spatial scale allows the models to learn and retain consistent feature representations. Note that most current global climate simulations and projections are also conducted at approximately $0.5^\circ$ resolution. Therefore, the models developed in this study can be more readily adapted, validated, or fine-tuned for application to other climate datasets without requiring substantial re-optimization. In this context, employing a uniform $0.5^\circ$ resolution for both training and testing increases the generalizability of our DL model and enhances the model applicability through transfer learning. This approach also mitigates systematic biases that could arise from training on a single dataset, allowing for a more robust assessment of the model’s ability to reconstruct historical TCG climatology.

Second, the focus of this study is on reconstructing TCG climatology, which is expected to be governed by environmental conditions at the synoptic scale rather than by inner-core dynamics at the early stages of TC development. Focusing on such large-scale environmental controls allows for a broader understanding of TCG processes from a climate perspective, especially in terms of how favorable large-scale conditions regulate TC formation. From this climate standpoint, a $0.5^\circ$ resolution is generally adequate for capturing the relevant environmental signals for large-scale TCG as demonstrated in previously studies \cite{Camargo_etal2007b, Cha_etal2020, Huang_etal2023, Vu_etal2021, NguyenKieu2024, Kieu_etal2023, Le_etal2025}. 

To label TCG events, we used the IBTrACS dataset \cite{IBTRACS}, which provides comprehensive global records of tropical cyclones. For the WNP basin, IBTrACS includes data from January 1890 through December 2023. An advantage of the IBTrACS dataset is that it is structured using standard synoptic time intervals, which align with those in both the ERA5 and MERRA-2 reanalysis datasets. This temporal consistency facilitates the integration of IBTrACS with reanalysis data, enabling an efficient and coherent framework for supervised DL development. 


As part of the process of constructing the binary TCG dataset from the IBTrACS data, it is important to note that TCs can form at any time of day, whereas IBTrACS provides storm reports at fixed 6-hour intervals. Consequently, the recorded genesis location in the best-track dataset may not precisely coincide with the actual position of TC formation. The extent of this discrepancy typically depends on the storm’s translational speed and the accuracy of vortex center detection algorithms. In this study, we use a data enrichment method to account for such uncertainty with an assumption that the TCG location does not vary significantly within a time window. This assumption allows us to have more positive TCG event, regardless of defining TCG as the first IBTrACS time entry or when the storm is first classified as a tropical depression (TD) \cite{KieuNguyen2024, Le_etal2025}. By default, all positive TCG samples are assigned a binary label of 1, while non-TCG (negative) samples are labeled as 0 for our training.

\subsection*{TCG data enrichment}
A key step of our DL workflow for reconstructing TCG climatology is the data-enrichment strategy. Unlike previous studies that used satellite observations to train DL models for classifying TCG-related cloud features or for short-range forecasts, reconstructing TCG climatology from climate reanalysis data poses a very different challenge. This is because (1) satellite data are unavailable prior to 1980 for use during inference, and (2) reanalysis datasets contain far less detailed information than high-resolution satellite imagery. As discussed in a recent study by \cite{NguyenKieu2024}, the ability of a DL model to extract TCG-relevant signals from reanalysis data is highly sensitive to domain selection, meteorological input variables, data resolution, or prediction lead time. Therefore, a robust strategy to enrich TCG information during training is essential for reconstructing TCG climatology using reanalysis data alone. 

Specific for the TCG study, we need to generate a sufficient training dataset that has a binary structure indicating a yes or no TCG event for DL training. The procedure of creating such a binary dataset turns out to be critical for our problem herein, as one needs to have not only a positive TCG dataset but also a proper negative set such that DL models can maximally learn the difference \cite{KieuNguyen2024, Le_etal2025}. On the one hand, the better the contrast between the positive and negative samples, the more likely DL models can pick up the key features that help improve the performance of ML models. On the other hand, the design of positive/negative samples must also ensure the practical applications in climate research and significance. For example, too selective negative samples would result in exceptionally well performance, but it also eliminates the model's generalization in applications.  

In this study, we first define a TCG reference as the first time a storm is classified as a TD, which is then labeled as a positive TCG event. This is a stage that is often highly identifiable on satellite images or global model products, thus reliably indicating the existence of a TC. One could also define a TCG as the first moment that a TC record is reported to make sure to capture the very first moment that a TC forms. However, our choice of the first time that a TC was recorded as a TD in the best track herein has the benefit of generating a range of data enrichment as described below and so we used this definition as a reference for any poisitive TCG event in the best track. 

To enrich the TCG dataset that can account for the uncertainty in TCG timing, we then selected all positive TCG labels and extended their timestamp backward by up to 48 hours, under the assumption that precursory signals of TCG may be present in the environments even prior to the TD classification. Apparently, there is some uncertainty regarding how far back such signals can be considered representative of a developing TC. This uncertainty is more pronounced during the pre-satellite era, due to the scarcity of reliable observations over open ocean. In this study, we approached this problem by using DL to conduct a series of sensitivity tests with different data enrichment windows to evaluate how far back TCG-related signals could still be meaningfully detected prior to the first TD record in IBTrACS. As shown in L25, a 48-hour window generally provides a sufficient balance among signal retention, label reliability, recall and precision scores. Thus, the 48-h window was used in this study. 

With this enrichment strategies, we scanned through all TC track records in the best-track database to create a set of TCG events. Given the typical scale of TCG, a TCG domain was chosen to be a square box of size 18$\times$18 degree, centered on the recorded TCG location for which reanalysis data was extracted (which is roughly $33\times32$ data points with the ERA5 or \merra\ resolution). Finally, all relevant information related to TCG events including their longitudes, latitudes, date, and time were stored in a csv database to facilitate our data sharing and input to the DL interface.


For the negative-labeled TCG data, one can have several different ways to define a negative TCG event, depending on the context and application purpose \cite{KieuNguyen2024, Le_etal2025}. In this study, we chose a simple approach of using the same location as the positive TCG-labeled data, but the data is taken back in time, starting from the last positive TCG label enrichment and ending after 5 days (see Fig. \ref{fig:resnet}a). This negative TCG strategy answers a question of why a TC forms within a time window but not earlier. For each positive TCG label, this strategy can therefore produce a series of negative TCG labels in the past, depending on how far back one wants to sample the negative labels and the data enrichment window.

With those negative TCG labels generated from both ERA5 and MERRA-2, we could effectively train our ResNet-18 model. There are several other important considerations regarding the adjustment of data ratios that must be taken into account to address the highly imbalanced nature of the dataset as well as the model's sensitivity to these adjustments. These data sensitivity analyses and implementation details specific to the ResNet-18 architecture are documented in L25 and so they are not repeated here.  

\subsection*{Deep learning model}
Rapid progress in DL research has constantly introduced new DL architectures for applications across research and practical fields. Specifically for TC, DL applications have surged recently, with many different approaches and improvements for different problems \cite{Gao_etal2018, Giffard_etal2020, miller_etal2017, Weyn_etal2021, NguyenKieu2024}. 

In the context of TCG application, several different approaches have been proposed to examine the skill of DL models in predicting TCG, which can be grouped into two types. The first type utilizes a set of TCG predictors to train a range of ML classification models. The second type uses spatial-temporal architectures to process image data such as satellite imagery, radar observation, or climate reanalysis to identify cloud or signal precursors to classify and track systems that are likely to develop into TCs \cite{Matsuoka2018, NguyenKieu2024}. This latter type is more computationally expensive, but it can account for the spatial inhomogeneity of environments during TCG. Thus, it can produce promising results despite limited training data, especially on the aspect of detecting and analyzing environmental patterns needed for TCG from input data. 

\begin{figure}[ht]
\begin{center}
\includegraphics[width=13cm]{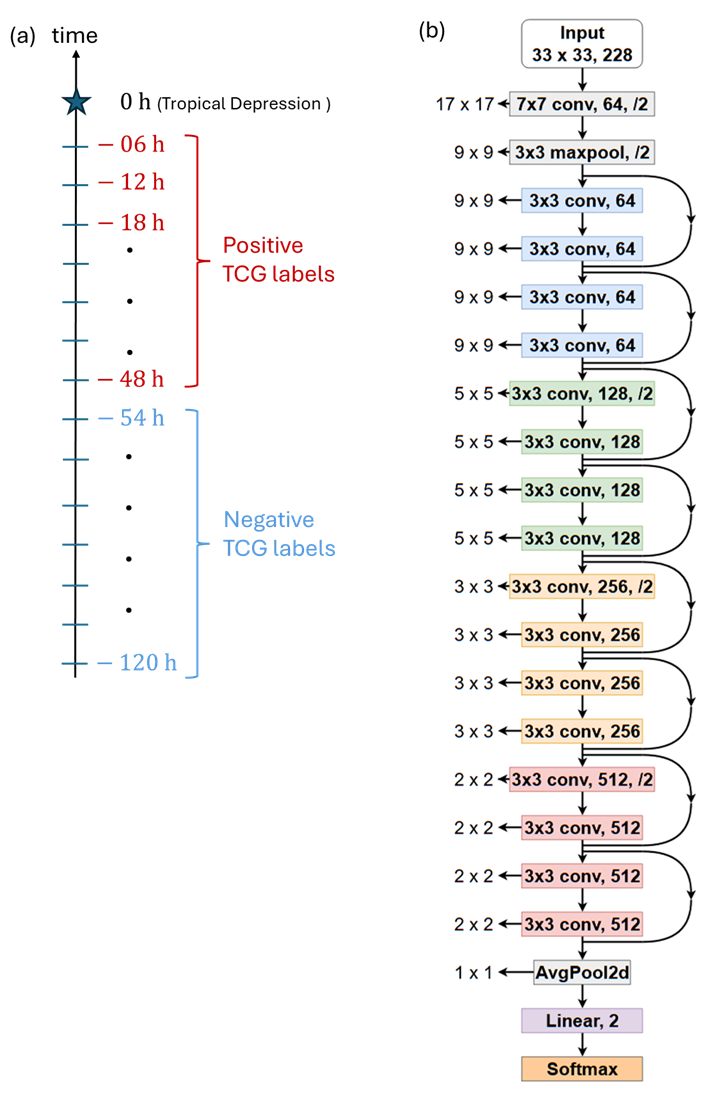}
\caption{a) A data enrichment design to generate TCG labels needed for training the ResNet-18 model, where the asterisk denotes the first timestamp in the best-track data that indicates the existence of a Tropical Depression (TD) stage; and b) the DL model design based on the ResNet-18 architecture that is used for TCG reconstruction in this study (adopted from \cite{Le_etal2025}).}
\label{fig:resnet}
\end{center}
\end{figure}

Because of such a capability of spatiotemporal architectures as well as the need for handling four-dimensional data volumes in climate reanalysis datasets, we adopted an 18-layer residual convolutional neural network (ResNet-18) model as in L25 for our TCG climatology reconstruction in this study. ResNet is a widely used architecture for image recognition tasks, whose key advantage is the use of residual blocks to avoid the gradient vanishing issue when the number of convolutional layers is sufficiently large (He et al. 2015). Using the same design as in L25, our DL model consists of a series of convolutional blocks, each has two convolutional layers, a batch normalization layer, and a ReLU activation function, along with a "shortcut" connection that directly adds the block's input to its output (Fig. \ref{fig:resnet}b). This design is effective in mitigating the "vanishing gradient" problem as shown in numerous image processing applications. More details of this ResNet-18 model and model sensitivity to different hyperparameters for reconstructing TCG climatology can be found in L25.  

In this study, we note that our implementation differs slightly from that in \cite{NguyenKieu2024} and \cite{Le_etal2025} in the use of full-level MERRA-2 for pre-training and then ERA5 data for model fine-tuning. Unlike the NCEP reanalysis used in \cite{NguyenKieu2024} or MERRA-2 data used in L25, our use of both the MERRA-2 and ERA5 datasets in this study do not include certain common environmental features such as tropopause-related variables. Thus, the ResNet-18 architecture employed here required additional refinement to accommodate the different input features for both datasets. As a result, the optimal model performance reported in this study differs from that in \cite{Le_etal2025}, despite a similar DL design.        
%
%
\clearpage
\clearpage

While our DL design may not rival the most recent state-of-the-art architectures in other problems, our experiments with several modern models such as vision transformers or graph neural networks did not outperform the simple ResNet-18 configuration used in this study in terms of TCG reconstruction. This likely reflects the nature of TCG reconstruction from 0.5$^\circ$-resolution climate data, which may offer limited additional information for more complex models to exploit given current datasets. For this reason, we do not expect our ResNet-18 model for reconstructing TCG climatology to be used directly for other TC-related metrics such as intensity, rapid intensification, accumulated cyclone energy, or the power dissipation index. These metrics depend not only on large-scale environmental conditions but also on fine-scale physical processes, which are still poorly represented in existing reanalysis products. This limitation underscores a broader challenge for DL-based approaches in practical TC applications that different TC metrics may require tailored modeling strategies or specialized architectures to achieve robust performance.



\subsection*{Model inference}
With the ResNet-18 model fully trained, the inference procedure for reconstructing the TCG map for any test period proceeds as follows. First, the entire WNP basin is partitioned into windows of fixed size $18^\circ\times18^\circ$. These windows are then slid across the basin in both the zonal and meridional directions using a $5^\circ$ increment. For each window, the DL model is applied to produce a TCG probability, which is assigned to the center of that window. This process is repeated over the entire test period at every 6-hour interval, consistent with the temporal resolution of the MERRA-2 or ERA5 datasets. Because the sliding interval is $5^\circ$, the resulting TCG reconstruction map has an effective spatial resolution of $5^\circ$, as shown in Figs. \ref{fig:dist_1722}–\ref{fig:dist_4060}. 

For the monthly distribution of TCG frequency, the procedure is similar to TCG density, except that the DL model is applied to each month separately. We then average all TCG probabilities over the entire domain obtained for that month, which is finally collected for display as seen in, e.g., Fig. \ref{fig:dist_monthly}. 

\bibliography{reference1, reference2, reference3}

\section*{Acknowledgments}
CK is partially supported by research grants from NSF (AGS-2309929). TN, TL, DH, LL, QL, BD, and TN are supported by the Vingroup Innovation Foundation (VINIF, code VINIF.2023.DA019).

\section*{Author contributions statement}
CK conceived the ideas and drafted the manuscript. TN, TL, KM, BD, and TD contributed to the overall deep-learning model development and operational workflow, DH, QL, and TN contributed to the implementation of the DL model system, data preprocessing, and conducted experiments. All authors contributed to reviewing, analyzing the results, and editing the manuscript. 



\end{document}